\documentclass[prd,twocolumn,superscriptaddress,showpacs,nofootinbib,floatfix,preprintnumbers]{revtex4}
\usepackage{axodraw}
\usepackage{graphicx,bm}

\def\beq {\begin{equation}}
\def\eeq {\end{equation}}
\def\bea {\begin{eqnarray}}
\def\eea {\end{eqnarray}}

\def \met  {\mbox{${E\!\!\!\!/_T}$}}

\newcommand{\br}{\begin{eqnarray}}
\newcommand{\er}{\end{eqnarray}}
\newcommand{\be}{\begin{equation}}
\newcommand{\ee}{\end{equation}}

\usepackage{color}
\usepackage{hyperref} 
\hypersetup{linktocpage=true}
\usepackage[all]{hypcap}
\hypersetup{
   bookmarks=true,         
   unicode=false,          
   pdftoolbar=true,        
   pdfmenubar=true,        
   pdffitwindow=false,     
   pdfstartview={FitH},    
   pdftitle={My title},    
   pdfauthor={Author},     
   pdfsubject={Subject},   
   pdfcreator={Creator},   
   pdfproducer={Producer}, 
   pdfkeywords={keyword1} {key2} {key3}, 
   pdfnewwindow=true,      
   colorlinks=true,       
   linkcolor=blue,          
   citecolor=magenta,        
   filecolor=magenta,      
   urlcolor=cyan           
}
\usepackage{fancyhdr}
\begin{document}
\preprint{HRI-RECAPP-2016-009} 
\title{Probing the Heavy Neutrinos of Inverse Seesaw Model at the LHeC}

\author{Subhadeep Mondal } \affiliation{Regional Centre for Accelerator-based Particle Physics,\\
Harish-Chandra Research Institute, Jhunsi, Allahabad - 211019, India}

\author{Santosh Kumar Rai} \affiliation{Regional Centre for Accelerator-based Particle Physics,\\
Harish-Chandra Research Institute, Jhunsi, Allahabad - 211019, India}


\vskip 0.3cm 
\begin{abstract}
We consider the production of a heavy neutrino and its possible signals at the Large Hadron-electron 
Collider (LHeC) in the context of an inverse-seesaw model for neutrino mass generation. 
The inverse seesaw model extends the Standard Model (SM) 
particle content by adding two neutral singlet fermions for each lepton generation. It is a well 
motivated model in the context of generating non-zero neutrino masses and mixings. The 
proposed future LHeC machine presents us with a particularly interesting possibility to probe 
such extensions of the SM with new leptons due to the presence of an electron beam in the initial state. 
We show that the LHeC will be able to probe an inverse scenario with much better efficacy compared
to the LHC with very nominal integrated luminosities as well as exploit the advantage of having the electron 
beam polarized to enhance the heavy neutrino production rates.
\end{abstract}
\maketitle
\section{Introduction}
The discovery of the Higgs boson \cite{Higgs:lhc} has been a remarkable achievement by the 
experiments running at the Large Hadron Collider (LHC), which in a profound way give closure to the 
predictions within the Standard Model (SM) picture of particle physics. However, some unanswered questions 
remain which forces us to look beyond the SM (BSM). One of the intriguing issues that needs immediate 
attention is the existence of tiny non-zero neutrino masses. Neutrino oscillation data reveals that at least 
two of the three light neutrinos of the SM are massive and also indicates a significant mixing among all the 
three neutrino states (for a review, see e.g, \cite{neut:osc}). SM, devoid 
of any right-handed neutrinos, fails to account for this particular phenomenological aspect. The 
simplemost natural extension of the SM comes in the form of type-I seesaw mechanism \cite{csaw} which 
adds one additional heavy Majorana neutrino to the SM particle content. As a consequence of mixing 
with this heavy neutrino state, one of the light neutrino gets a non-zero mass at the tree level. 
Another neutrino gains non-zero mass at one loop level, reproducing the neutrino oscillation data 
perfectly. However, the smallness of the neutrino masses ensure that either the left-right Yukawa coupling 
is very small, $\sim 10^{-6}$ or the heavy Majorana neutrino mass is extremely heavy, $\sim 10^{16}$ GeV, 
making the new features of this model virtually untestable at the present collider experiments. 

Inverse Seesaw \cite{Mohapatra:1986aw,Nandi:1985uh,Mohapatra:1986bd} is a well motivated BSM scenario 
from the  viewpoint of neutrino mass generation. Owing to the presence of a very small lepton number 
violating parameter, $\mu_S\sim$ eV which is responsible for the smallness of the light neutrino masses in 
the model, the Yukawa coupling generating the Dirac neutrino mass term can be quite large ($\sim$ 0.1) even 
in the presence of sub TeV heavy neutrino masses in this scenario. This leads to a plethora of 
phenomenological implications in 
non-supersymmetric \cite{Das:2012ze, Bandyopadhyay:2012px,Dev:2013wba,Das:2014jxa,Arganda:2014dta,Deppisch:2015qwa,
Arganda:2015naa,Arganda:2015ija,Das:2015toa} 
as well as in supersymmetric context 
\cite{Hirsch:2009ra,Mondal:2012jv,BhupalDev:2012ru,DeRomeri:2012qd,Banerjee:2013fga}. 
Two obvious aspects of any neutrino mass model which can then be probed at the collider experiments are, 
production of the additional heavy neutrino states and studying the effects of left-right neutrino mixing. As 
the Yukawa couplings are large in the inverse seesaw model, it is possible to obtain flavour specific leptonic 
final states through their production and subsequent decays. 
Search for such heavy neutrinos in this kind of models, at LHC has been discussed in detail 
\cite{Datta:1993nm,Han:2006ip,delAguila:2007qnc,Huitu:2008gf,Atre:2009rg,Das:2012ze,Dev:2013wba,Das:2014jxa,Chen:2011hc,
Alva:2014gxa,Deppisch:2015qwa,Das:2015toa}. 
Electron enriched final states are usually suppressed in the usual models of neutrino mass generation 
due to the stringent constraints derived from the non-observation of neutrinoless double beta decay 
($0\nu\beta\beta$) \cite{Deppisch:2015qwa} on the heavy neutrino mixing with $\nu_e$. 
In the inverse seesaw mechanism, however, this constraint is relaxed due to the extremely small 
mass splitting 
between the heavy neutrinos with opposite CP-properties, forming a quasi-Dirac state owing to a 
${\mathcal O} (eV)$ lepton number violating (LNV) parameter, $\mu_S$. 
As a matter of fact, the smallness of $\mu_S$ forces all the 
LNV processes to be suppressed in such scenarios. Therefore, the usual smoking gun LNV signals 
of heavy Majorana neutrino are not ideal to look for in inverse seesaw model. However, 
a loose $0\nu\beta\beta$ constraint opens the additional possibility for an electron enriched 
final state. 
Heavy neutrino production associated with an electron has been studied at the LHC  in the context 
of an inverse seesaw model \cite{Das:2012ze}. It was observed that one expects 5$\sigma$ statistical 
significance over the SM backgrounds for a trilepton signal at the 14 TeV run of LHC with 
11~${\rm fb}^{-1}$ integrated luminosity and degenerate heavy neutrino masses of 100 GeV.  
It turns out that the trilepton signal is by far the best channel to probe the inverse seesaw model 
at the LHC. However, if the heavy neutrino states are much more massive, then these heavier states 
will be produced with much reduced rates which in turn affects the sensitivity of probing 
these heavy states at the LHC. An alternative search strategy for much heavier states can, in fact 
be carried out more efficiently at a different scattering experiment such as the proposed Large 
Hadron-Electron Collider (LHeC)  \cite{AbelleiraFernandez:2012cc,Bruening:2013bga} which is the 
main thrust of this work. To highlight this, we study the heavy neutrino production in the inverse seesaw 
model at LHeC and determine its signal strengths through various leptonic channels. 
   
LHeC would be the next high energy $e-p$ collider after HERA, supposed to be built at the LHC tunnel. 
The design is planned so as to collide an electron beam with a typical energy range, 60-150 GeV with 
a 7 TeV proton beam producing center of mass energy close to 1.3 TeV at the parton level. 
It is expected to achieve 100~${\rm fb}^{-1}$ integrated luminosity per year. Lepton number violating 
heavy Majorana neutrino signals and other phenomenological consequences have already been explored in the context of the 
$e-p$ colliders \cite{Ingelman:1993ve, Liang:2010gm,Blaksley:2011ey,Duarte:2014zea,Mondal:2015zba,Lindner:2016lxq} 
and future lepton colliders \cite{Basso:2013jka,Antusch:2014woa,Antusch:2015mia}. 
Note however that LHeC has a distinct advantage over the LHC for this kind of searches since the electron 
in the initial state can be polarised. This very interesting and important aspect of using the 
polarisation of the initial electron beam to study specific BSM scenarios at LHeC was first pointed out 
by us in Ref.\cite{Mondal:2015zba}. 
A dominantly {\it polarised} (left/right) electron beam could thereby enhance 
a new physics signal for specific production channels while also affecting the corresponding SM 
background, making it an additional tool to explore BSM physics as was envisaged for linear 
electron-positron colliders. 
Although for a high energy ($\sim 100$ GeV) electron beam it is difficult to maintain a high enough 
polarisation, for a 60 GeV beam, polarisation of upto 80$\%$ can be 
achieved \cite{AbelleiraFernandez:2012cc}.  Such a machine will therefore help determine quite 
distinctly the nature of specific production modes of the heavy neutrinos \cite{Mondal:2015zba}.    
\section{The Model}

The SM particle content is extended by the addition of two fermion singlets to each generation. 
These two singlets, $N^c$ and $S$ are assigned lepton numbers -1 and +1 respectively. The extended 
Lagrangian looks like:
\begin{eqnarray}
	{\cal L}=\epsilon_{ab}y_\nu^{ij} L^a_i H^b N^c_j + {M_R}^{ij}N^c_i S_j + {\mu_S}^{ij} S_i S_j,
\label{lag-invs}
\end{eqnarray}   
where $\mu_S$ is a small ($\sim {\rm eV}$) lepton number violating ($\Delta L = 2$) parameter. 
The $9\times 9$ neutrino mass matrix in the basis $\{\nu_L, N^c, S\}$ 
looks like:
\begin{eqnarray}
	{\cal M}_\nu = \left(\begin{array}{ccc}
		{\bf 0} & M_D & {\bf 0}\\
		M_D^T & {\bf 0} & M_R \\
		{\bf 0} & M_R^T & \mu_S
	\end{array}\right),
	\label{eq:mbig}
\end{eqnarray}
where, $M_D =  y_{\nu} \, v$ is the Dirac neutrino mass matrix, $v\simeq 174$ GeV being the vacuum 
expectation value (vev) of the Higgs field in the SM. Under the approximation, $\|\mu_S\|\ll \|M_R\|$ 
(where $\|M\|\equiv \sqrt{{\rm Tr}(M^\dag M)}$), one can extract the $3\times 3$ light neutrino mass matrix. 
Upto leading order in $\mu_S$ it looks like: 
\begin{eqnarray}
	M_\nu = \left[M_DM_R^{T^{-1}}\right]\mu_S\left[(M_R^{-1})M_D^T\right]\equiv F\mu_S F^T ,
	\label{eq:vmass}
\end{eqnarray}
where $F=M_DM_R^{T^{-1}}$.
It is evident from Eq.~(\ref{eq:vmass}), that the smallness of neutrino mass here depends on the smallness 
of the lepton-number violating parameter $\mu_S$ instead of the smallness of $M_D$ and/or heaviness of 
$M_R$ as in the canonical type-I seesaw case. Consequently, one can have a $M_R$ below the ${\rm TeV}$ 
range even with a comparatively large Dirac Yukawa coupling, ($y_{\nu}\sim 0.1$). Both $M_D$ and $M_R$ 
can be kept strictly diagonal \footnote{A diagonal $M_D$ is also favoured from the strictly constrained 
lepton flavour violating (LFV) decay branching 
ratios, which get enhanced rapidly in case it has non-negligible off-diagonal entries.} and the neutrino 
oscillation data can be fit by an off-diagonal $\mu_S$, where,  
\begin{eqnarray}
\mu_S=F^{-1}M_\nu {F^T}^{-1}
\label{eq:mus-fit}
\end{eqnarray}
$M_\nu$ can easily be constructed from the neutrino oscillation parameters :
\begin{eqnarray} 
M_\nu=U_{PMNS}M_{\nu}^{\rm diag.}U^T_{PMNS},
\label{eq:mnu-rec}  
\end{eqnarray}
where, $M_{\nu}^{\rm diag.}$ is the diagonal neutrino mass matrix and $U_{PMNS}$ is the diagonalising unitary 
Pontecorvo-Maki-Nakagawa-Sakata (PMNS) matrix. In order to construct the $U_{PMNS}$ and $M_{\nu}^{\rm diag.}$ 
matrices, We have considered the most updated neutrino oscillation parameters \cite{Gonzalez-Garcia:2015qrr} obtained from global fit of the experimental data. The oscillation parameters for normal hierarchy in neutrino masses as well as the 
resulting PMNS matrix we used are presented in Table~\ref{tab:osc}.
\begin{table}[h!]
\begin{center}
\begin{tabular}{||c|c||} 
\hline 
Parameters & Values   \\
\hline\hline  
${\rm sin}^2 \theta_{12}$ & $0.304^{+0.013}_{-0.012}$ \\
\hline
${\rm sin}^2 \theta_{23}$ & $0.452^{+0.052}_{-0.028}$ \\
\hline
${\rm sin}^2 \theta_{13}$ & $0.0218^{+0.001}_{-0.001}$ \\
\hline
$\Delta m^2_{21}~{\rm eV}^2$  & $(7.50^{+0.19}_{-0.17})\times 10^{-5}$ \\
\hline
$\Delta m^2_{32}~{\rm eV}^2$ & $(2.457^{+0.047}_{-0.047})\times 10^{-3}$ \\
\hline
$U_{PMNS}$ & {\tiny $\left(\begin{array}{ccc}
0.825 & 0.545 & 0.148\\
-0.491 & 0.563 & 0.665\\
0.280 & -0.621 & 0.732\\
\end{array}\right)$} \\
\hline\hline 
\end{tabular}
\caption{Three flavor neutrino oscillation data obtained from global fit for normal hierarchy in neutrino masses
as presented in Ref.~\cite{Gonzalez-Garcia:2015qrr} and the resulting PMNS matrix.}
\label{tab:osc} 
\end{center}
\end{table} 

Once we have $M_{\nu}$ in the form of Eq.~\ref{eq:mnu-rec}, it can be plugged back into 
Eq.~\ref{eq:mus-fit} to find  the $\mu_S$ matrix. 
\section{Heavy neutrino signals}

Same sign dilepton signature has been studied extensively in order to probe the possible Majorana nature 
of heavy neutrinos and their mixing with the light SM-like neutrinos. At the LHC, the usual channel 
that is considered is $pp\rightarrow W^{\pm}\rightarrow\ell^{\pm}N$, where $N$ subsequently decays into a 
lepton associated with on-shell or off-shell $W$ which then decays hadronically. For Majorana 
neutrinos, this results in the same-sign dilepton final state ($\ell^{\pm}\ell^{\pm}jj$) with 
negligible missing energy. 
However, cross-section for this final state is rendered small either by the constrained Yukawa coupling 
or heavy Majorana neutrino masses. Most stringent constraint on these parameters in case of usual 
neutrino mass models, e.g. type-I seesaw extended SM, is derived from the non-observation of 
neutrinoless double beta decay 
($0\nu\beta\beta$); $\sum_N \frac{V_{eN}}{m_N} < 5\times 10^{-8}~{\rm GeV}^{-1}$ \cite{Deppisch:2015qwa}. 
Thus, heavy Majorana neutrinos which couple to the electron are usually neglected at collider experiments
\footnote{However, if the light-heavy mixing can be rendered small, as can be done in presence of some 
extended symmetry group \cite{Gluza:2015goa,Gluza:2016qqv}, the $0\nu\beta\beta$ constraint may be evaded resulting in interesting 
phenomenological consequences with the first generation leptons.}.
However, in the case of inverse seesaw extended SM, the contribution to $0\nu\beta\beta$ is extremely 
small due to the presence of a small $\mu_S$ as already mentioned. Hence this particular constraint 
becomes non-restrictive for electron channels \cite{Das:2012ze}. Therefore it becomes an imperative 
channel to probe an inverse seesaw scenario at a machine like LHeC where the electron-type 
heavy neutrino can be directly produced.  
Non-observation of any Majorana neutrino signal at the LHC so far from di-muon final state translates 
into a bound on the light-heavy neutrino mixing parameter, $|V_{\mu N}|^2\sim 10^{-2}-10^{-1}$ for 
heavy neutrino masses $m_N$=100-300 GeV \cite{ATLAS:2012ak,ATLAS:2012yoa,Chatrchyan:2012fla}. 

The special feature of an inverse seesaw extended SM is that it can simultaneously have a $\sim$ 100 GeV 
heavy Majorana neutrino with $\sim$ 0.1 Yukawa coupling even after satisfying neutrino oscillation data. 
This makes it particularly interesting phenomenologically. In this article, we look into the various possible 
final states in this framework in the context of LHeC. The fact that the LHeC can produce an electron beam 
that can be highly polarised upto a certain energy, makes it even more interesting phenomenologically to 
probe such scenarios. Since the SM singlet neutrino only mixes with the left-handed leptons, a 
dominantly left-polarised electron beam in the initial state is expected to increase the 
cross-section significantly.  Additionally, being an $e-p$ collider, LHeC will produce much cleaner 
signals compared to the LHC.
\subsection{Analysis}

Here we intend to look into the possible production modes of the heavy neutrino.  The dominant 
contribution comes from the channel $e \, p \to N \, j$ while another (sub-dominant) contributing 
process can be identified as $e \, p \to N \, j \, W^-$, where $N$ indicates the heavy 
right-handed neutrino states. 
An inclusive contribution to the signal also  comes from a heavy neutrino mediated process 
given by $e \, p \to e^{-} \, j \, W^{-}$.  This may result into final states consisting of different 
lepton-jet multiplicity and missing energy. 

\begin{figure}[h!]
\centering
\includegraphics[height=4.0cm,width=8cm]{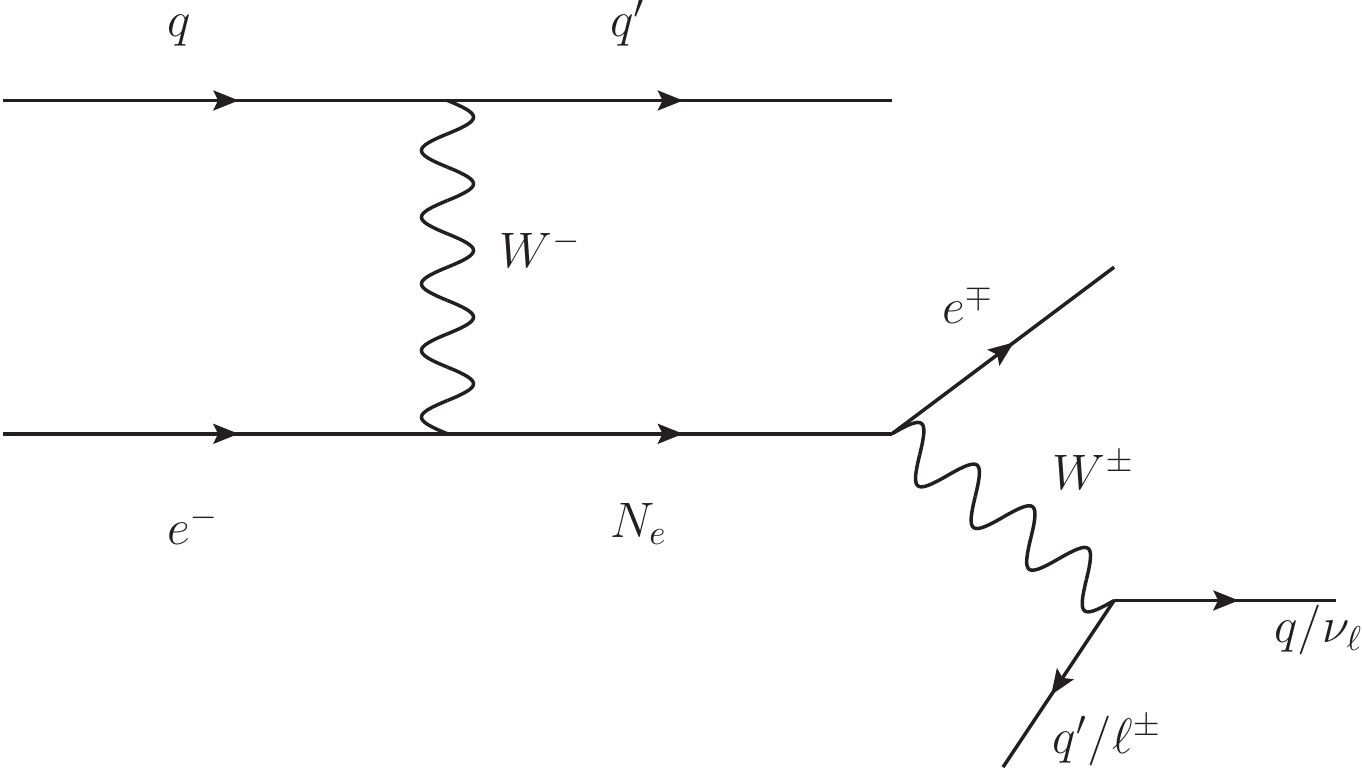}
\caption{Feynman diagram for the various final states derived from $Nj$ production channel.}
\label{fig:feyn1}
\end{figure}

\begin{figure}[h!]
\includegraphics[height=4cm,width=8cm]{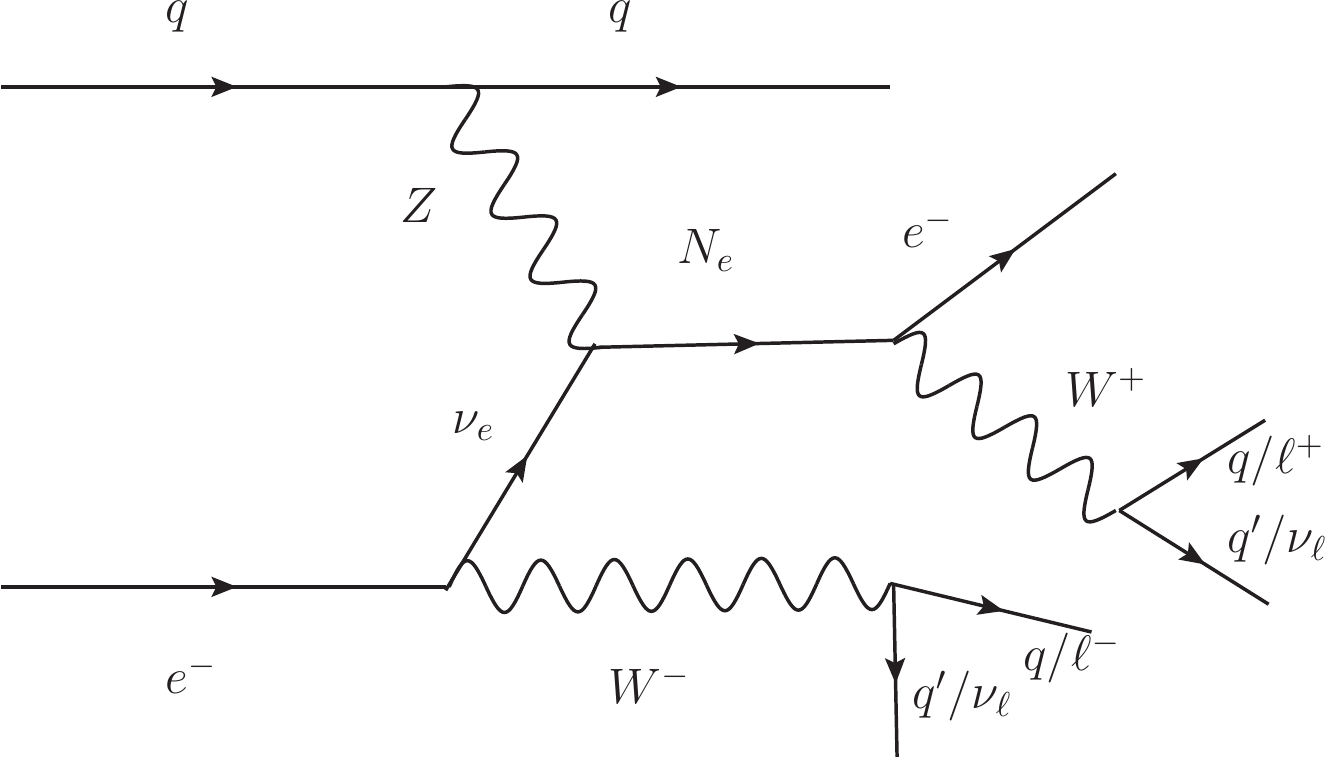}
\caption{Feynman diagram for the various final states derived from $NjW^-$ production channel.}
\label{fig:feyn2}
\end{figure}

\begin{figure}[h!]
\centering
\includegraphics[height=4cm,width=8cm]{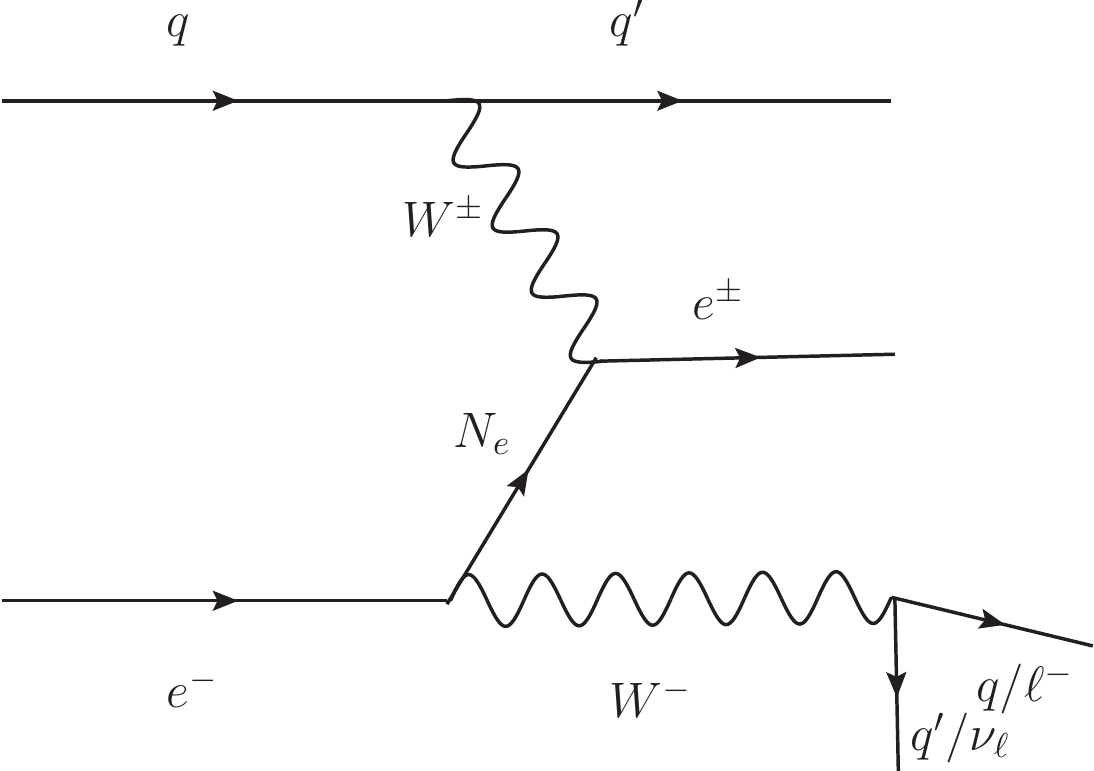}
\caption{Feynman diagram for the various final states derived from $e^{-}jW^{-}$ production channel.}
\label{fig:feyn3}
\end{figure}

In Fig.~\ref{fig:feyn1} we show the $Nj$ production channel and subsequent decays of $N$  
to give rise to the following possible final states (at parton-level): 
\begin{itemize}
  \item $e^+$/$e^-$ + n-jets ($n=3$)
  \item $e^{\pm}\ell^{\mp}$ + n-jets ($n=1$) + $\met$.
\end{itemize}
Here, $\ell$ represents either an electron or muon in the final state. 
This clearly gives the most dominant contribution to the signal for a heavy neutrino production.
However, a higher jet- and lepton-multiplicity signal also can arise from another production mode for $N$, 
which is usually neglected but can in principle also contribute to the signal rates arising from $Nj$ 
production. This is shown in Fig.~\ref{fig:feyn2}, where the $NjW^-$ production channel and 
subsequent decays of $N$ and $W^-$, give rise to the following possible final states (at parton-level): 
\begin{itemize}
  \item $e^+$/$e^-$ + n-jets ($n=5$)
  \item $e^{\mp}\ell^-$ + n-jets ($n=3$) + $\met$
  \item $e^{\mp}\ell^{\pm}\ell^-$ + n-jets ($n=1$) + $\met$.
\end{itemize}  
As one can see that all the channels listed for the $NjW^-$ mode can in principle contribute to the 
final states arising from the $Nj$ production mode, where the higher multiplicity in jets or leptons can 
be reduced through mismeasurements, trigger efficiencies, etc.  Finally, an $N$ mediated channel 
($e^{-}jW^{-}$), which is at the same order in coupling as the $NjW^-$ process, may also contribute 
to the signal which we therefore include in our analysis. This subprocess is shown in Fig.~\ref{fig:feyn3} 
and gives rise to the following possible final states (at parton-level) which is exactly the same 
as $Nj$ production: 
\begin{itemize}
  \item $e^{\pm}$ + n-jets ($n=3$)
  \item $e^{\pm}\ell^{-}$ + n-jets ($n=1$) + $\met$.
\end{itemize}
Note that, although we show all the possible final states arising from the above three production channels, 
including the lepton number violating processes which have extremely small cross-sections 
within the framework of inverse seesaw model, as already mentioned.
Hence among all the above listed final states we shall ignore the LNV contributions and only 
concentrate on the lepton number conserving ones, which arise in the inverse seesaw scenario.    

\begin{figure}[h!]
\centering
\includegraphics[height=9cm,width=9cm,angle=270]{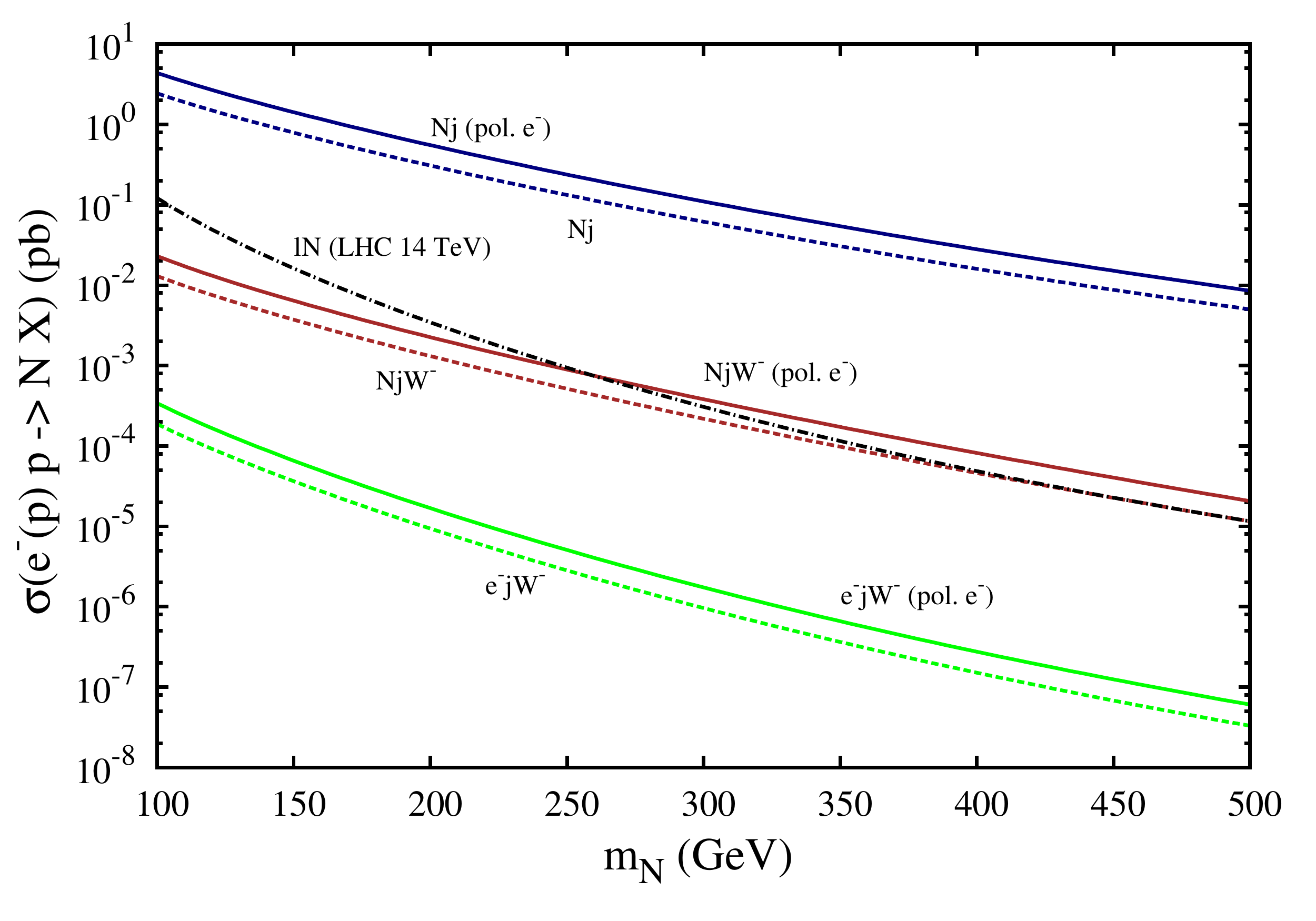}
\caption{Production cross-section of $Nj$, $NjW^-$ and $e^-jW^-$ at LHeC with and 
without 80\% left-polarised electron beam along with that of the most dominant production 
mode of the heavy neutrinos at the LHC, $\ell N$, with varying heavy neutrino masses. 
The blue solid and dotted lines represent the variation of production cross-section in $Nj$ channel 
with and without the left-polarised electron beam. Likewise, the brown and the green lines represent 
the $NjW^-$ and $e^-jW^-$ production channels respectively.  The black (dashed) line corresponds 
to the $\ell N$ production cross-section at the LHC at 14 TeV center of mass energy.}
\label{fig:xsec}
\end{figure}

We now discuss the production rates in the different channels both in the context 
of LHC and LHeC. While calculating the cross-sections at LHC, we chose the center of mass energy 
of 14 TeV which would give the largest rate for LHC. Note that, at the LHC, both $p p\to N j$ and 
$p p \to N j W^-$ production channels will be absent since both of them are lepton number violating 
processes ($\Delta L=1$), whereas models for Majorana neutrinos allow only $\Delta L=2$ violations. 
Therefore the dominant production channel at LHC for the heavy neutrino is  
$pp\rightarrow W^{\pm}\rightarrow\ell^{\pm}N$. For LHeC, we have considered a 60-GeV electron 
beam colliding with a 7 TeV proton beam. We compute the cross-section with both 
polarised\footnote{Note that, when we mention polarised electron beam , we mean an electron beam 
that is 80\% left-polarised.} and unpolarised electron beams to assess how much the cross-section may 
actually differ. 
We calculate the cross section for the different production modes of the heavy neutrino as a 
function of its mass, shown in Fig.~\ref{fig:xsec}. We have parametrized the heavy-light neutrino 
mixing term as a function of the heavy neutrino mass, using the condition $V_{eN}=y_\nu^{11} \, v/m_N$ 
where we have fixed $y_\nu^{11}=0.1$. Note that this gives a substantially conservative estimate of the
mixing for heavier masses, as the limits for such mixing is expected to be weaker as the mass $m_N$ 
increases.
    
Fig.~\ref{fig:xsec} shows the variation of the cross-section of the aforementioned production channels
at LHeC. As a comparison to what LHC may achieve in terms of production rates for such a 
heavy neutrino, we also show the variation of the cross-section, $\sigma(pp\to\ell N)$ which is
the most dominant production channel of the heavy neutrino at the LHC \footnote{A competitive 
production channel takes over this production mode for heavier neutrino masses above 
$\sim 600$ GeV \cite{Dev:2013wba,Alva:2014gxa,Degrande:2016aje}.}.  
The blue, brown and green dotted lines correspond to the cross-sections at LHeC with an unpolarised 
electron beam for $Nj$, $NjW^-$ and $e^-jW^-$ production modes respectively. The corresponding solid 
lines of same colour show the relative enhancement in the cross-sections when the electron beam is 
dominantly left-polarised (80\%). 
As one can see, the rates improve by almost a factor of 2 over the entire range of the heavy neutrino 
mass in case of a polarised $e^-$ beam. 
The cross-section for $\ell N$ production at the 14 TeV run of LHC, shown in black (dashed) 
line in Fig.~\ref{fig:xsec}, is quite small when compared to the $Nj$ production mode at LHeC. 
It however does compete in the low $m_N$ region with the LHeC production rates of the 
significantly sub-dominant $NjW^-$ production. Interestingly, for heavier masses, LHeC with a 
polarised electron beam provides a better rate even in this mode. Note that the $e^-jW^-$ channel which 
was included is unlikely to be of any significance due to its extremely small 
cross-section. The large suppression, when compared to the $NjW^-$ production can be understood
from the fact that the proton radiates a $W^-$ in the t-channel for $e^-jW^-$  while a $Z$ boson
mediates the $NjW^-$. This implies that while the valence $u$ and $d$ quarks of the proton contribute 
to the $NjW^-$ production, $e^-jW^-$ production is only through the valence $d$ quark. Thus, 
we can safely ignore this production channel for the rest of our analysis.

We choose to carry out our analysis for two particular benchmark points to probe one of the several heavy 
neutrinos in the model. It is evident from the neutrino mass matrix in Eq.~\ref{eq:mbig} that $\mu_S$ being 
a very small parameter, the masses of the heavy neutrino states are dictated by the choice of the matrix 
$M_R$.  Among the six heavy neutrino states, there exist three mass degenerate pairs. We 
chose to keep the lightest of these heavy neutrino pairs at sub-TeV mass and make the rest of them 
very heavy. For example, we chose  ${M_R}^{11} \simeq  m_N$=150 GeV while fixing 
${M_R}^{22}$=${M_R}^{33}$=1000 GeV. For simplicity, a diagonal structure for both $M_R$ and $y_{\nu}$ 
is assumed.  As a consequence, the $m_N=150$ GeV heavy neutrino states will only couple to the electron. 
Accordingly we have fit the neutrino oscillation data with an off-diagonal $\mu_S$ which we found can
easily accommodate all experimental results of the neutrino sector.  Similarly, another benchmark is 
set with $m_N=400$ GeV.  For our collider analysis, we keep the mixing $V_{eN}$ for this benchmark to 
be the same as that for $m_N=150$ GeV. Table~\ref{tab:bps} shows our choices of the neutrino sector 
parameters in order to fit the neutrino oscillation data. 
\begin{table}[h!]
\begin{center}
\begin{tabular}{||c|c|c||} 
\hline 
Parameters & BP1 & BP2  \\
\hline\hline  
$M_R$ (GeV) & (150.0,1000.0,1000.0) & (400.0,1000.0,1000.0) \\
\hline
$y_{\nu}$ & (0.1,0.01,0.01) & (0.1,0.01,0.01) \\
\hline
$\mu_S$ (keV) & {\tiny $\left(\begin{array}{ccc}
0.0003 & 0.0380 & 0.0126\\
0.0380 & 8.3424 & 7.1666\\
0.0126 & 7.1666 & 10.1186\\
\end{array}\right)$} & {\tiny $\left(\begin{array}{ccc}
0.0020 & 0.1014 & 0.0335\\
0.1014 & 8.3424 & 7.1666 \\
0.0335 & 7.1666 & 10.1186  \\
\end{array}\right)$} \\
\hline\hline 
\end{tabular}
\caption{Choices of the neutrino sector input parameters $M_R$ and $y_\nu$ and the resulting $\mu_S$ 
matrix from neutrino oscillation data for our two benchmark points. Note that, in order to fit 
the neutrino oscillation data, we consider the experimental constraints assuming only 
normal hierarchy among the neutrino masses. The corresponding experimental constraints are 
taken from Ref.~\cite{Gonzalez-Garcia:2015qrr}.}
\label{tab:bps} 
\end{center}
\end{table} 
With these choices of the parameters, we now proceed to 
study the signal arising from the production channels $NjW$ and $Nj$. 

The complete model has been implemented in {\tt SARAH} (v4.6.0) \cite{sarah}. The mass spectrum 
for the inverse seesaw model has been subsequently generated using {\tt SPheno} (v3.3.6) 
\cite{spheno} along with the mixing matrices and decay strengths. 
We have used {\tt MadGraph5@aMCNLO} (v2.3.3) \cite{mad5} to generate parton level events for 
the signal and backgrounds. Subsequent decay, showering and hadronisation are done using 
{\tt PYTHIA} (v6.4.28) \cite{pythia6} whereas a proper {\tt MLM} matching procedure has been 
employed for the multi-jet final states. 

To analyse the signal and background events we use the following set of selection criteria to isolate 
the leptons and jets in the final state:
\begin{itemize}
\item Electrons and muons in the final state should have $p_T^{\ell} > 20$ GeV,$|\eta^{e}| < 2.5$ and 
$|\eta^{\mu}| < 2.5$.    
\item Photons are counted if $p_T^{\gamma} > 10$ GeV and $|\eta^{\gamma}| < 2.5$ as the leptons.
\item Leptons should be separated by, $\Delta R_{\ell\ell} > 0.2$. 
\item Leptons and photons should be separated by, $\Delta R_{\ell\gamma} > 0.2$.
\item For the jets we require $p_T^{j} > 40$ GeV and $|\eta^j| < 2.5$.
\item Jets should be separated by, $\Delta R_{j j} > 0.5$.
\item Leptons and jets should be separated by, $\Delta R_{\ell j} > 0.4$.
\item Hadronic energy deposition around an isolated lepton must be limited to 
$\sum p_{T_{hadron}} < 0.2\times p_T^{\ell}$.
\end{itemize}  
As already mentioned, because of the choice of digonal $M_R$ and $y_{\nu}$, the $N$ always decays into 
an $e$ or $\nu_e$ associated final state. Here we only consider $N$ decaying into an electron and $W$. 
Hence all the possible final states consist of an electron accompanied by other leptons and jets. 
Although the lepton number violating final state $e^+$ + n-jets derived from both 
$NjW$ and $Nj$ production channels have practically negligible 
SM background\footnote{A recent proposal that electron charge-misidentification (cMID) can lead to 
a serious source of background is based on some presumptive choices of 
efficiencies, as large as 1\%  \cite{Lindner:2016lxq} . 
A recent CMS analysis \cite{Khachatryan:2016olu} on LNV signal at LHC clearly give estimates of the 
cMID efficiency (charge-flip) that range between $\sim 3.2 \times 10^{-5} - 2.4 \times 10^{-4}$, which 
practically renders this background at LHeC negligible. In addition, one expects the LHeC, with a cleaner 
environment,  would further improve on this cMID efficiency, putting much doubt that charge misidentification 
can possibly be a major source of background for LNV processes which one needs to worry about \cite{Queiroz:2016qmc,Mondal:2016czu}.}  
their signal cross-section is also highly suppressed due to the small lepton number violating 
parameter. Hence LNV final states are not ideal to probe inverse seesaw scenario. Instead, 
we concentrate only on the lepton number conserving final states arising from both the production 
channels.  Apart from the cascade, some additional jets are expected to appear from the initial 
and (or) final state radiation while showering. Hence we chose various final states consisting of a 
minimum number of jets that we expect to come from the cascades itself, depending upon the leptonic 
or hadronic decays of the $W$-boson. 

The SM background events were generated for all the subprocesses that contribute towards the different 
final states that we consider for the signal. We have explicitly chosen to compute the 
continuum SM background where the dominant contributions arise mostly from the gauge boson 
mediated $t$-channel processes.  After selecting both the signal and background events  through the 
aforementioned selection criteria, we further put a minimal $\met$ requirement of 20 GeV on all final states
under consideration. 

\subsection{Results}
\begin{figure}[h!]
\centering
\hspace{-1cm}
\includegraphics[scale=0.33,angle=270]{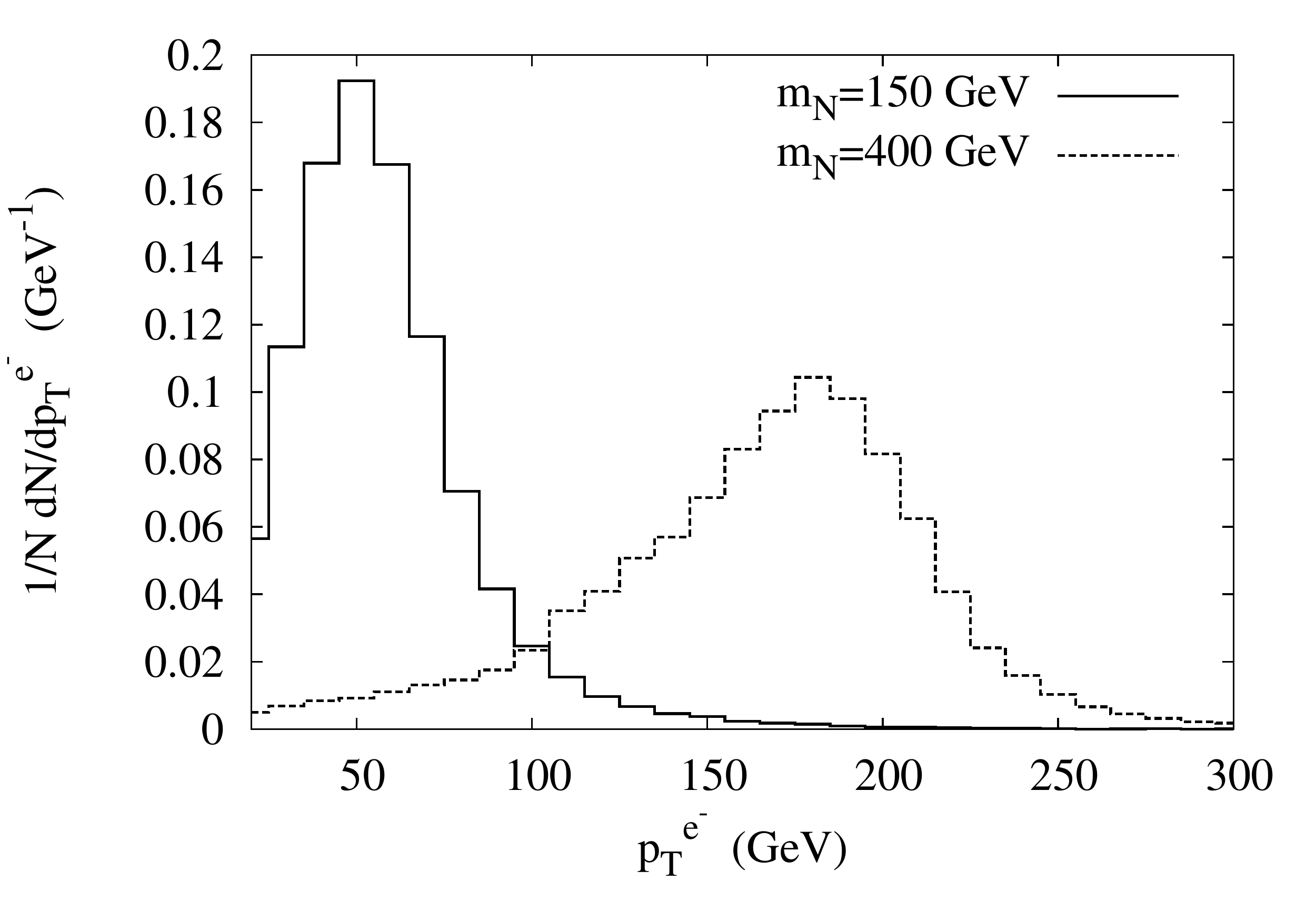}
\caption{Transverse momentum of the $e^-$ originating from decaying $N_e$ for $m_N=150$ GeV and 
400 GeV obtained from the $Nj$ production channel.}
\label{fig:elec_pt}
\end{figure}

In this section, we present the results of our simulation at two different heavy neutrino masses, 
$m_N=150$ GeV and 400 GeV. For the $m_N=150$ GeV case, we apply the aforementioned cuts 
whereas for the heavier neutrino mass scenario we impose slightly harder cuts to reduce 
SM backgrounds more effectively. Note that for the signal, the electron coming from the decay of 
the heavy neutrino would carry a $p_T$ depending on the mass difference $\Delta M=m_N-M_W$. 
Thus for a much heavier neutrino such as $m_N=400$ GeV, a stronger $p_T$ requirement on the 
leading $e^-$ would help. This is also evident from Fig.~\ref{fig:elec_pt}, which shows the above 
mentioned feature for our two different mass choices of $N$.
We therefore demand that the leading electron must have a $p_T>100$ GeV for the $m_N=400$ GeV
signal. We keep the rest of the selection criterion same as for the $m_N=150$ GeV.     
We present in Table~\ref{tab:cross-sec}, the resulting cross-section for all the 
signal channels and also for the corresponding SM backgrounds whereas in Table~\ref{tab:sig}, 
we show the required integrated luminosity to achieve a $3\sigma$ excess in the four 
final states we have studied at the LHeC. Note that as the $NjW^-$ production cross-section becomes 
negligibly small for $m_N=400$ GeV, we only present the final state contributions for this benchmark, 
arising from the $Nj$ production mode. 

\begin{table}[h!]
\begin{center}
\begin{tabular}{||c|c|c|c|c||} 
\hline 
$m_N$ & Final States & $\sigma$ (fb) & $\sigma$ (fb) & $\sigma$ (fb)  \\
(GeV) & & (NjW)  & (Nj) & (SM)  \\
\hline\hline  
150 & $e^-$ + n-jets ($n\geq 3$) + $\met$ & 0.36 & 12.86 & 38.05 \\
& $e^-\ell^-$ + n-jets ($n\geq 3$) + $\met$ & 0.02 & - & 0.01 \\
& $e^-\ell^+$ + n-jets ($n\geq 1$) + $\met$ & 0.68 & 87.68 & 24.45 \\
& $e^-\ell^+_i\ell^-_j$ + n-jets ($n\geq 1$) + $\met$ & 0.04 & - & 0.72 \\
\hline
400 & $e^-$ + n-jets ($n\geq 3$) + $\met$ & - & 1.42 & 15.21 \\
& $e^-\ell^+$ + n-jets ($n\geq 1$) + $\met$ & - & 2.20 & 6.30 \\
\hline\hline 
\end{tabular}
\caption{Signal cross-sections for various final states obtained from $NjW$ and $Nj$ productions 
computed for two different heavy neutrino masses. 
The last column represents corresponding background cross-sections 
as obtained from the SM. Note that, the quoted cross-sections are obtained for different sets of cuts 
at the two different heavy neutrino masses.}
\label{tab:cross-sec} 
\end{center}
\end{table} 
We calculate the statistical significance for the signal using  
\begin{eqnarray}
    {\mathcal S} = \sqrt{2 \times \left[ (s+b){\rm ln}(1+\frac{s}{b})-s\right]}, 
\end{eqnarray}
where, $s$ and $b$ represent the signal and background event counts respectively, 
by combining the event rates of both the production channels $Nj$ and $NjW^-$. 
\begin{table}[h]
\small
\begin{center}
\begin{tabular}{||c||c|c|}
\cline{1-3}
\multicolumn{1}{||c||}{}&
\multicolumn{2}{|c|}{Required luminosity  ($\mathcal{L}$) } \\
\multicolumn{1}{||c||}{}&
\multicolumn{2}{|c|}{for ${\mathcal S}=3{\sigma}$ (in $fb^{-1}$)} \\
\multicolumn{1}{||c||}{}&
\multicolumn{2}{|c|}{} \\
\cline{2-3}
\multicolumn{1}{||c||}{Final States}&
\multicolumn{2}{c|}{$m_N $  (in GeV)} \\
 & 
$150$ & $400$   \\ 
\cline{1-3}
$e^-$ + n-jets ($n\geq 3$) + $\met$ 
& 2.2 & $ 70.0 $\\ 
$e^-\ell^-$ + n-jets ($n\geq 3$) + $\met$
& 347.3 & - \\
$e^-\ell^+$ + n-jets ($n\geq 1$) + $\met$
& 0.05 &  13.0 \\
$e^-\ell^+_i\ell^-_j$ + n-jets ($n\geq 1$) + $\met$
& $4.12\times 10^{3}$ & - \\
\hline
\end{tabular}
\caption{An estimate of the required integrated luminosity to achieve a $3\sigma$ 
statistical significance for each of the four final states for two different heavy 
neutrino masses at LHeC.}
\label{tab:sig}
\end{center}
\end{table}

All the final states are associated with at least one $e^-$ originating from the $N$ decay, 
irrespective of how the $W$'s have decayed.  The rate of the final state 
$e^-$ + n-jets ($n\geq 3$), arising from $NjW^-$ production is rather small due to the small 
production cross-section, but non-vanishing for the $m_N=150$ GeV signal. 
For the heavier neutrino mass region, this contribution can be easily neglected. However, 
the combined rate of this final state is reasonably good with bulk of the contribution arising 
from the $Nj$ production. Overall, this final state proves to be a viable one to probe such a scenario 
at the LHeC even with low integrated luminosities. This fact is illustrated in Table~\ref{tab:sig}, 
which indicates that a $3\sigma$ statistical significance can be achieved with an 
integrated luminosity ($\mathcal{L}$) 
as low as  $\simeq 2 \,{\rm fb^{-1}}$ for the $m_N=150$ GeV case while 
a $5\sigma$ signal would require $\mathcal{L}\simeq 6\,{\rm fb^{-1}}$. For the 
$m_N=400$ GeV case, requirement of harder $p_T$ of the $e^-$ reduces the background 
contribution quite effectively. However, the signal suffers because of the smallness of the 
$Nj$ production cross-section itself. However, this benchmark too presents a signal in the excess of 
$3\sigma$ statistical significance with a slightly higher integrated luminosity of  
$\simeq 70 \,{\rm fb^{-1}}$, as shown in Table~\ref{tab:sig}, while a $5\sigma$ signal  
would require $\mathcal{L}\simeq 200\,{\rm fb^{-1}}$. Comparing this with what the LHC might 
be able to do for such a mass region, clearly emphasises how a machine such as LHeC can play 
a very crucial role in studying models of neutrino mass generation, surpassing LHC sensitivities in 
a very short span of time. 

If the $W^-$ in $NjW^-$ production mode decays leptonically, then it can lead to a same-sign 
dilepton final state associated with jets and $\met$. This final state is characteristic to the $NjW^-$ 
production channel, since the $Nj$ channel only can give rise to opposite-sign dileption final states. 
However, inspite of a much reduced SM background contribution, the same-sign dilepton final state 
is expected to be significant only at a relatively higher luminosity ($\sim 350~{\rm fb^{-1}}$) due to its 
small signal rate even at lighter right-handed neutrino masses.  
On the other hand, the opposite-sign dilepton 
final state turns out to be the most significant channel to search for the heavy neutrinos at lower mass range 
at the LHeC. Such final states may be obtained if $W^+$ originating from the decay of $N$, decays 
leptonically. As evident from Table~\ref{tab:cross-sec}, for $m_N=150$ GeV, the 
$e^-\ell^+$ + n-jets ($n\geq 1$) + $\met$ signal rate from $Nj$ production channel is larger than the 
SM background contribution for our choice of the benchmark point. The corresponding statistical significance 
is quite good indicating the fact that this channel is capable of 
probing much smaller values of light-heavy neutrino mixing ($|V_{eN}|$) as well as 
signals of much heavier neutrino mass scenarios. For example when $m_N=150$ GeV, one can probe 
$|V_{eN}| \sim 10^{-2}$ with a statistical significance of $3\sigma$ using 100 fb$^{-1}$ integrated 
luminosity, i.e. just an year of LHeC running. Note that for such a small value of mixing, the corresponding 
LHC reach at its 14 TeV run becomes practically negligible without the very-high luminosity option. Even for 
the much heavier $m_N=400$ GeV case, the required  integrated luminosity is reasonably 
small ($\mathcal{L}\simeq 13~{\rm fb^{-1}}$) for a  $3\sigma$ signal at the LHeC, while 
$\mathcal{L}\simeq 36~{\rm fb^{-1}}$ would give a $5\sigma$ discovery, which once again 
highlights the importance of such a machine.  
    
Finally, the trilepton final state may arise when the $W$'s in $NjW^-$ only decay leptonically. Such a 
final state cannot be obtained from the $Nj$ production mode and hence has a smaller event rate. 
It can, therefore, be an option only for the lower $m_N$ region. However, in the vicinity of 
$m_N=150$ GeV, although the signal cross-section is quite small, increased lepton multiplicity leads 
to significant reduction in SM background contribution. This channel, therefore, may be a good 
complimentary channel to the discovery channel at  very high luminosities ($\sim$ 4100~${\rm fb^{-1}}$).   

Note that LHeC also has a possibility of ramping up the electron energy to 150 GeV. This invariably leads 
to increased rates for both the signal and SM background. The increased energy option for the electron 
beam shows that the signal cross section for the $Nj$ production increases by a factor of $\sim 1.7$ for 
$m_N=150$ GeV while the rate improves by a factor of $\sim 3$ for $m_N=400$ GeV. Thus there can be 
further improvement in the sensitivity to heavier neutrino masses at LHeC.

\section{Conclusion}
To summarise, we have considered an inverse seesaw extended SM which is a very well motivated 
model from the viewpoint of the existence of non-zero neutrino masses and significant mixing among the 
neutrino states. This model is particularly phenomenologically interesting due to the possibility of having 
sub-TeV heavy neutrino states in the model alongwith a ${\mathcal O}$ (0.1) Yukawa coupling inducing 
significant left-right mixing in the neutrino states. We have explored the possiblity to detect the heavy neutrinos 
of this model in the context of LHeC. The presence of a left-polarized electron beam at the LHeC can enhance 
the production cross-section of the heavy neutrinos significantly.  We have looked at different possible 
final states arising from the heavy neutrino production through the $Nj$ and $NjW^-$ channels and their 
corresponding SM backgrounds. We conclude that LHeC with a polarised electron beam can be very 
effective to probe such models. We have also shown that even in the absence of a polarised electron beam, 
the $Nj$ production mode at the LHeC can be more effective to probe heavier right-handed neutrino masses 
or smaller neutrino Yukawa couplings than at the LHC. We observe that the $e^-\ell^+$ + n-jets ($n\geq 1$) 
+ $\met$ and $e^-$ + n-jets ($n\geq 3$) + $\met$ final states are the most promising discovery channels. 
In addition to that, for the lighter $m_N$ scenario, $e^-\ell^+_i\ell^-_j$ + n-jets ($n\geq 1$) + $\met$ may be a 
good complimentary channel at higher luminosities. However, one should note that, the LHeC snsitivity is 
limited to electron specific final states and can only probe smaller $|V_{eN}|$ than LHC. $|V_{\mu_N}|$ may 
be probed only if sizable mixing is allowed between the right-handed neutrinos associated with the first two 
leptonic generations. However, such mixing is highly constrained from non-observation 
of any lepton flavor violating decays. The flavour specific final states in such scenarios can help to predict 
the degree of flavour violation, if allowed, in such models.     

\bigskip
\begin{acknowledgments}
This work was partially supported by funding available from the Department of Atomic Energy, 
Government of India, for the Regional Centre for Accelerator-based Particle Physics (RECAPP), 
Harish-Chandra Research Institute.
\end{acknowledgments}

\end{document}